\documentclass{iaus}
\usepackage{graphicx}

\title[High velocity LMC and formation of the MS] 
{Modeling a high velocity LMC. \\ The formation of the Magellanic Stream.}

\author[Chiara Mastropietro]   
{Chiara Mastropietro$^{1,2}$}

\affiliation{$^1$LERMA, Observatoire de Paris, UPMC, CNRS,\\
61, A. de l'Observatoire, 75014, Paris, France, \\ 
email: {\tt chiara.mastropietro@obspm.fr}\\[\affilskip]
$^2$Universit\"ats Sternwarte M\"unchen, \\
Scheinerstr.1, D-81679 M\"unchen, Germany}

\pubyear{2008}
\volume{256}  
\pagerange{119--126}
\jname{The Magellanic System: Stars, Gas, and Galaxies}
\editors{Jacco Th. van Loon \& Joana M. Oliveira, eds.}
\begin{document}

\maketitle

\begin{abstract}
I use high resolution N-body/SPH simulations to model the new proper
motion of the Large Magellanic Cloud (LMC) within the Milky Way (MW)  halo  
and investigate the effects of gravitational and
hydrodynamical forces on the formation of the
Magellanic Stream (MS).
Both the LMC and the MW are fully self consistent galaxy models
embedded in extended cuspy $\Lambda$CDM dark matter halos.\\
I find that ram-pressure from a low density ionized halo is sufficient
to remove a large amount of gas from the LMC's disk forming a trailing
Stream that extends more than 120 degrees from the Cloud.
Tidal forces elongate the satellite's disk but do not affect its vertical
structure. No stars become unbound showing that tidal
stripping is almost effectiveless.

\end{abstract}

\firstsection 
\section{Introduction}
Recent HST proper motion measurements of the Magellanic Clouds 
by Kallivayalil et al. (20006a,b)  and \cite{ Piatek07} indicate that they  
are presently moving at 
velocities substantially
higher (almost 100 km s$^{-1}$) than those provided by previous
observational studies (van der Marel et al. 2002, Kroupa  \& Bastian 1997). 
Such high velocities ($v=378$ km s$^{-1}$ and  $v=302$ km s$^{-1}$ for
the LMC and SMC, respectively) 
are close to the escape velocity of the Milky Way
and consistent with the hypothesis 
of a first passage about the Galaxy (Besla et al. 2007).
A single perigalactic passage  has serious
implications for the origin of the Magellanic Stream.
It definitely rules out the tidal stripping hypothesis (Ruzicka, Theis \&
Palous 2008) since in this scenario
the loss of mass is primarily induced by tidal shocks suffered by satellites at the
pericenters (Mayer et al. 2006)  and the Stream would not have time to form before the
present time. Indeed kinematical data suggest that the Clouds are now just after a
perigalactic passage.
On the other hand, ram-pressure scales as $v^2$, where $v$ is
the relative velocity between  satellites and the ambient medium.
The high velocities of the Clouds could therefore  compensate the effect of
the reduced interaction time with the hot halo of the MW and
hydrodynamical forces would play a determinant role in forming the
Stream.\\
In \cite{Mastropietro05} (hereafter M05) we have performed high resolution N-body/SPH simulations to
study the hydrodynamical and gravitational interaction between the LMC
and the MW using orbital constraints by \cite{vanderMareletal02} and a
present time satellite velocity of 250 km s$^{-1}$. 
We found that, after two perigalactic passages, the combined effect of tidal forces and
ram-pressure stripping can account for the majority of the LMC's
internal features and for the formation of the MS. 
More in detail, ram-pressure stripping of cold gas from the LMC's disk
produces a Stream with morphology and kinematics similar to the
observed ones, while tidal stripping has
longer time-scales and is not efficient in forming stellar debris,
consistently with the lack of stars observed in the Stream.
Nevertheless, at each pericentric passage the LMC suffers tidal
heating which perturbs the overall structure of the satellite
reducing the gravitational restoring force and therefore indirectly contributing
to the loss of gas.\\
The main objection to this model, in light of the new proper
motion measurements of the LMC,  is that
the time spent by the LMC within the hot halo of the MW would be to short to cover the full
extension of the Stream (more than 100 degree) by ram-pressure mechanisms (Besla et al. 2007).
Moreover, hydrodynamical forces would affect a galaxy only weakly perturbed by the gravitational interaction and stripping would result more difficult.

In this work I present the results of N-body/SPH simulations
where the interaction between the MW and the LMC is modeled according
to the new proper motion measurements of \cite{Kallivayaliletal06a}.

\section{Galaxy models}

The initial condition of the simulations are constructed using the
technique described by \cite{Hernquist93}. Both the MW and the LMC are
multi-component galaxy models with a stellar and gaseous disk embedded
in a spherical dark matter halo. The density profile of the NFW halo
is adiabatically contracted due to baryonic cooling. Stars and cold
gas in the disks follow the same exponential surface density profile. We also
explore the eventuality of an extended LMC gaseous disk. In this model
the gaseous disk is characterized by an additional  
constant density layer which extends up to
eight times the scale length of the exponential disk.
The MW model comprises also a small stellar bulge and an extended low
density ($n = 2 \times 10^{-5}$ cm$^{-3}$ within 150 kpc from the Galactic center and $n = 8.5 \times 10^{-5}$ cm$^{-3}$ at 50 kpc ) hot ($T=10^6$ K) halo in hydrostatic equilibrium inside the
Galactic potential (M05).
The MW model, with virial mass $10^{12} M_{\odot}$ and concentration $c=11$,  is similar to model A1 of \cite{Klypin02} while  the 
structural parameters of the LMC are chosen in such a way that the resulting rotation curve
resemble that of a typical bulgeless late-type disk galaxy.
In details, the satellite has virial mass $2.6 \times 10^{10}M_{\odot}$, concentration $c=9.5$ and the same amount of mass in the stellar and gaseous disk component ($\sim 10^9 M_{\odot}$).
The Toomre's stability criterion is always satisfied and the parameter $Q$  set equal to  1.5 and 2.0 at the disk scale radius in the different LMC models.\\

\section{Simulations}

I performed adiabatic simulations  using GASOLINE, a parallel tree-code 
with multi-stepping (Wadsley et al. 2004). 
High resolution runs have $2.46 \times 10^6$ particles, of which
$3.5 \times 10^5$ are used for the disks and $5 \times 10^5$ for the
hot halo of the MW. The gravitational spline softening is set equal to
0.5 kpc for the dark and gaseous halos, and to 0.1 kpc for stars and
gas in the disk and bulge components.

 \begin{figure}[h!]
 \includegraphics[width=2.7in]{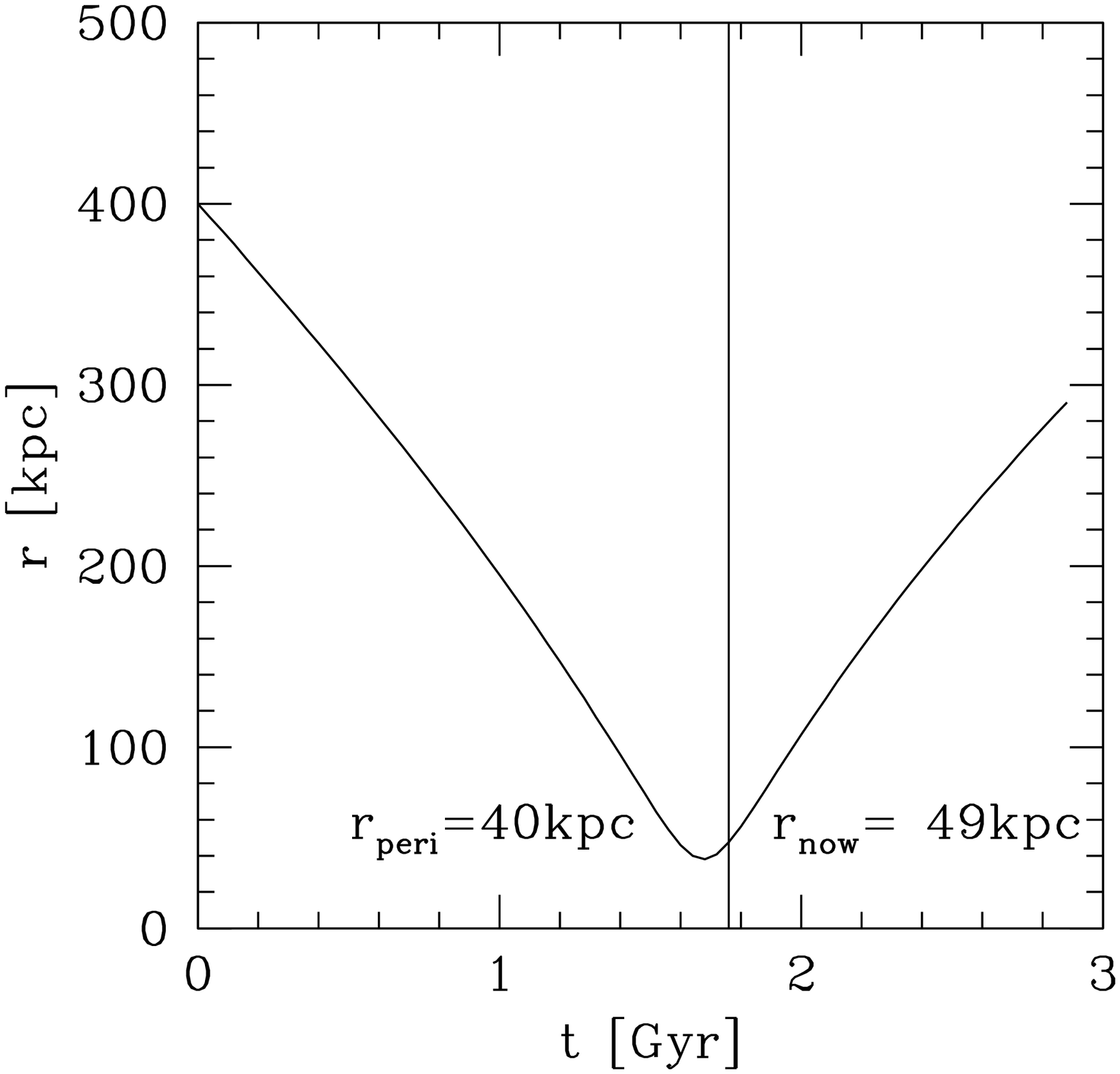} 
 \includegraphics[width=2.7in]{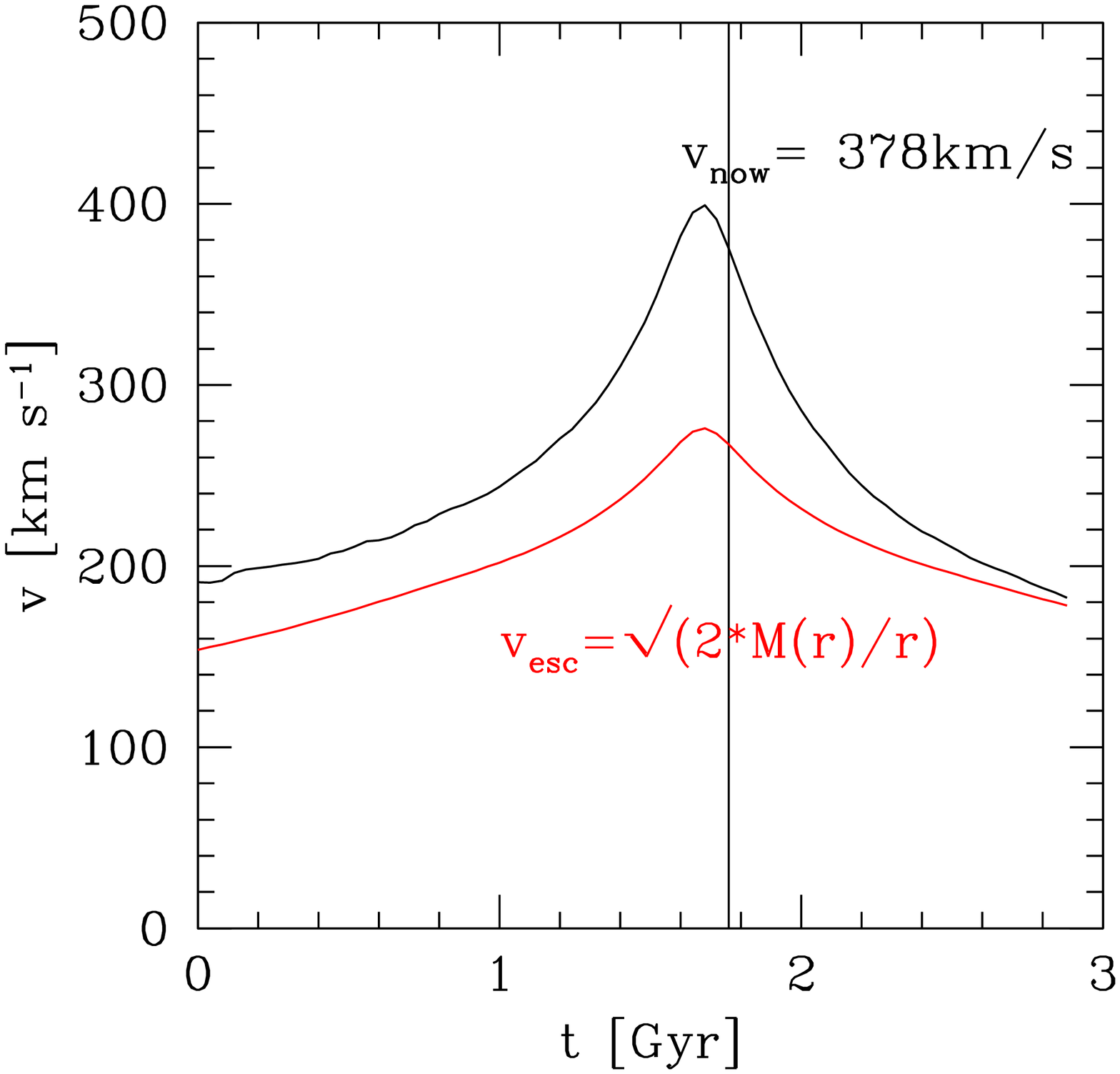} 
 \caption{Orbit (left panel) and orbital velocity (right) of the LMC. 
 Vertical lines represents the present time, just after the
 perigalacticon. Present values and values at the perigalacticon are indicated.   }
   \label{orbit}
\end{figure}

In my best model the LMC approaches the MW on an unbound orbit
 (Fig. \ref{orbit}) with initial Galactocentric distance  of 400 kpc and velocity $\sim 190$ km s$^{-1}$.
After
the perigalactic passage (at $\sim 40$ kpc) the velocity decreases faster than for a
ballistic orbit as a result of dynamical friction.
The escape velocity at a given LMC position is indicated by a red curve in the right panel and calculated assuming a
spherical unperturbed host potential.
Due to the effects of dynamical friction, at late times 
the satellite lies on a nearly parabolic orbit.
At the present time ($t \sim 1.78 $ Gyr, vertical lines in the plots)
it reaches a velocity of $378$ km $s^{-1}$ at $ 49$ kpc from the Galactic
center, in good agreement with the new proper motion measurements.\\
 In choosing the initial inclination of the LMC I made the
approximation that it does not change during the interaction due to
the effects precession or nutation of the disk plane.
At the beginning of the simulation the disk moves almost face-on through the external medium, 
and ram-pressure affecting the whole disk perpendicularly. In proximity to the perigalactic passage the velocity vector changes rapidly and the angle between the satellite's disk and the proper motion is close to zero.
At the present time the simulated disk has an inclination of about 30 degree 
with respect to the orbital motion (Kallivayalil et al. 2006a) and is indeed moving nearly edge-on through the external hot gas, with ram-pressure compressing its eastern side.

\begin{figure}[h!]
\vspace*{ 1.8 cm}
\begin{center}
 \hspace*{-1.0 cm}
 \includegraphics[%
  scale=0.8]{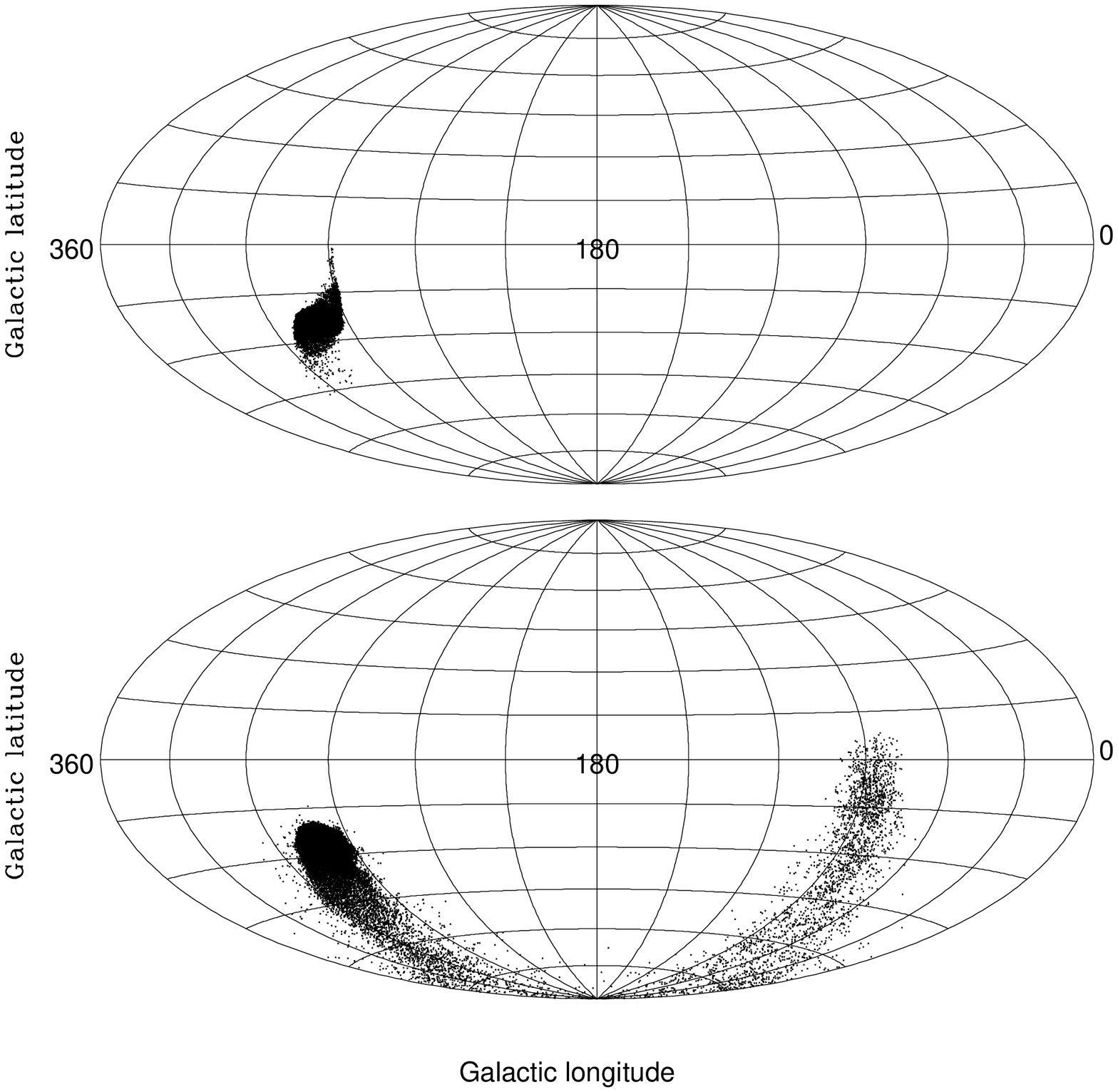}  
 \vspace*{ 1.5 cm}
 \caption{Present time distribution of stars (top) and gas (bottom)
 from the LMC disk in Galactic coordinates.}
   \label{aitoff}
\end{center}
\end{figure}

\begin{figure}[h!]
\vspace*{0.5 cm}
\begin{center}
\hspace*{3.5 cm}
 \includegraphics[%
  scale=0.6]{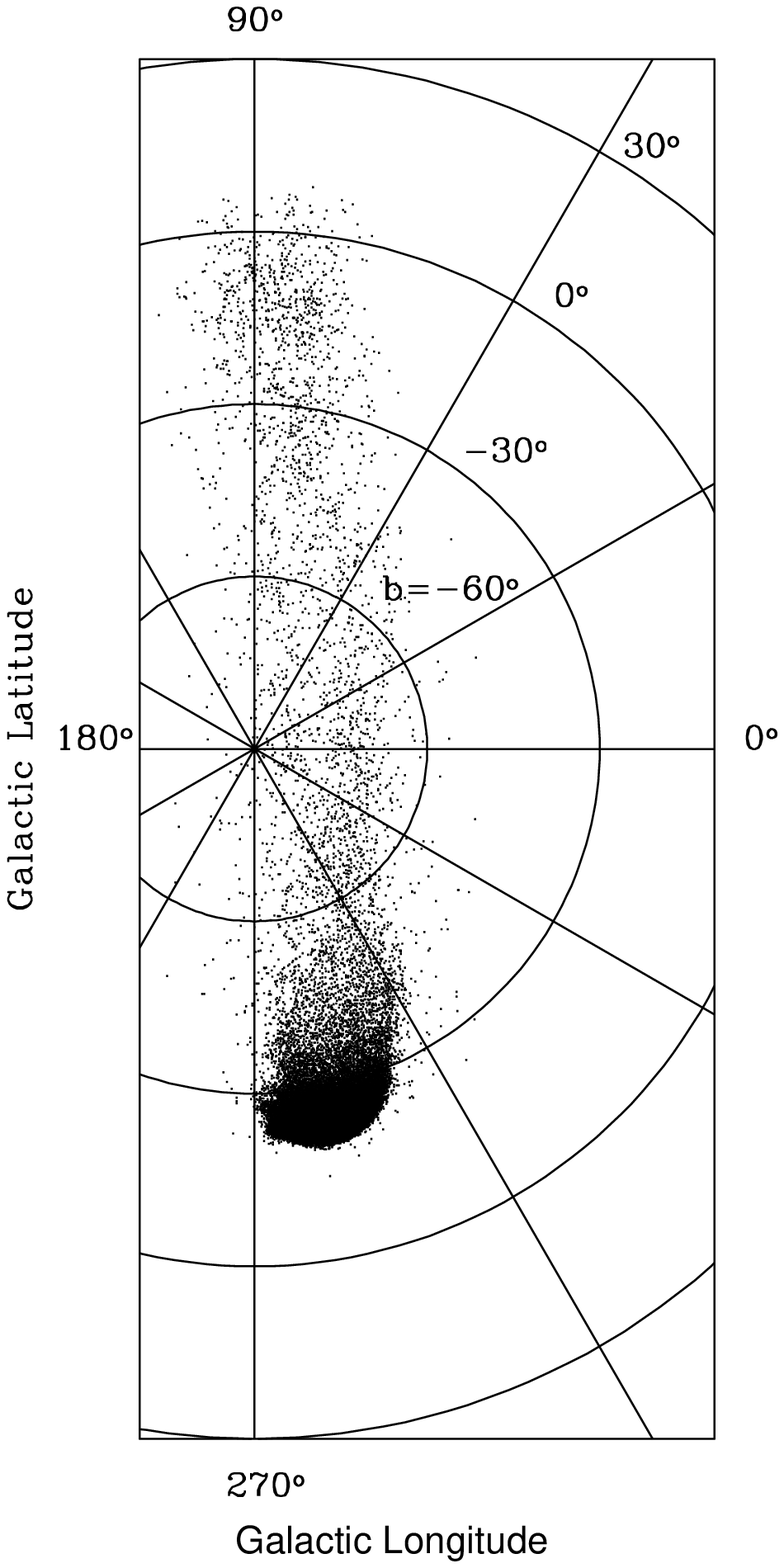} 
\hspace*{3.5 cm}
 \includegraphics[%
  scale=0.6] {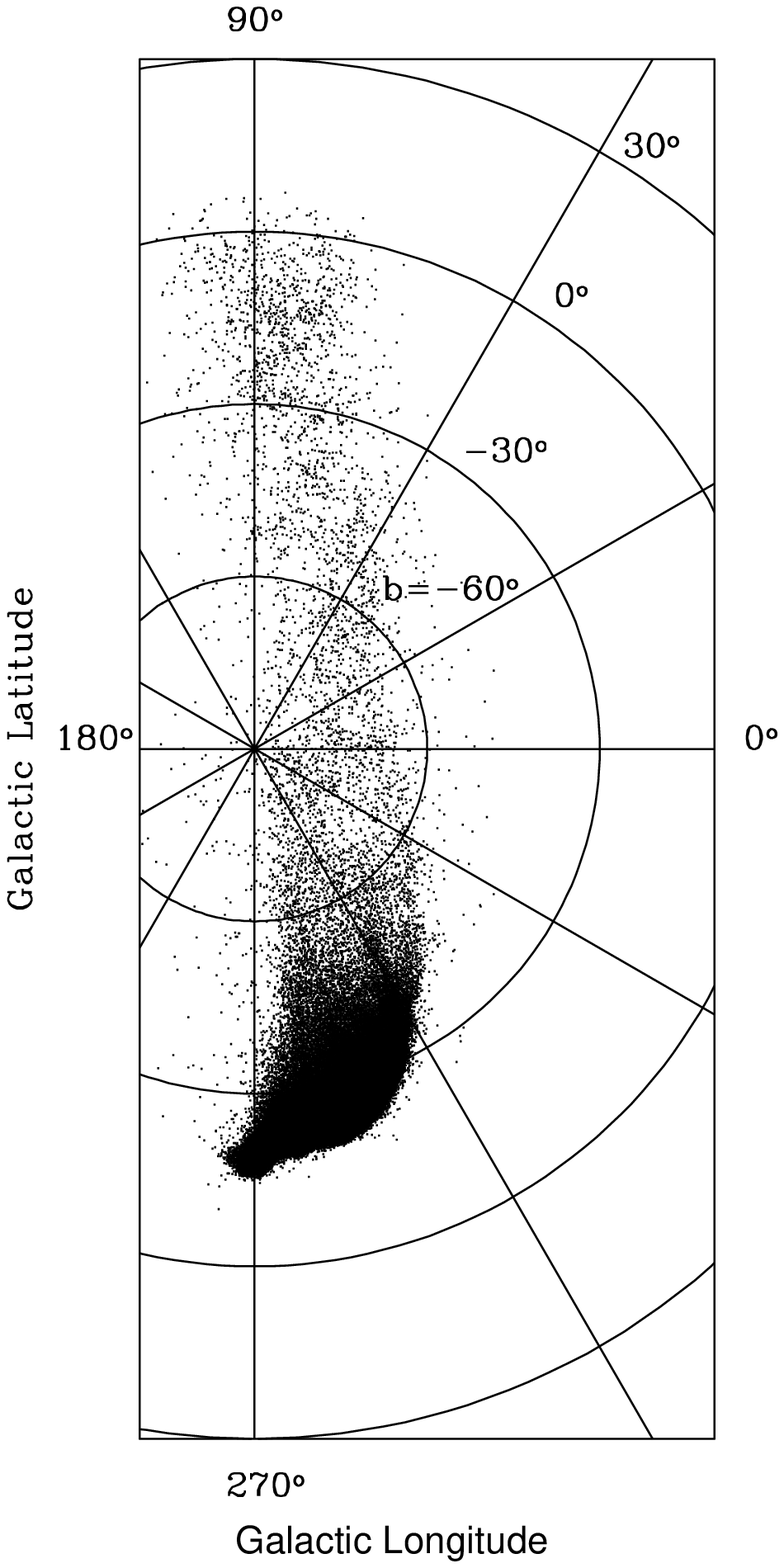} 
\vspace*{0.5 cm}
 \caption{Polar projection of the simulated stream in Galactic
   coordinates. Both the pure exponential LMC model (top panel) and
   the model with extended disk (bottom) are shown.
}
   \label{aitoffcirc}
\end{center}
\end{figure}

Fig. \ref{aitoff} illustrates the present time distribution of  stars and
gas originating from the LMC's disk. The stellar disk becomes elongated while
 tidal debris start forming after the perigalactic passage. But all  stars
stay bound to the satellite.
Tidal heating does not perturb significantly the vertical structure
of the disk that remains thin and does not create a warp, unlike 
what observed in M05.
Bar instability develops at the perigalacticon only in the case of $Q=1.5$.\\
Ram-pressure strips  nearly $2 \times 10^{8}
M_{\odot}$ of gas from the LMC's disk forming a continuous 
Stream that lies in a thin plane perpendicular to the disk of the MW and
extending up to $\sim 140 $ degrees from the LMC
(Fig. \ref{aitoffcirc}).
The location of the Stream in the Southern Galactic hemisphere is
comparable to the values of $b$ and $l$ provided by observations.
Contrary to M05 there is no LMC's gas above the
Galactic plane. In M05 the material lying in the
Northern hemisphere is stripped from the satellite during the 
orbital period preceding the present one and the  Stream forms a great polar circle
around the Galaxy.
The lack of gas at $b>0$ in Fig. \ref{aitoffcirc} is not due to
inefficient ram-pressure at early times, but to
the fact that, in order to reproduce the current location and velocity
of the LMC, the satellite enters the MW halo exactly at $b=0$. 

The morphology of the Stream does not change significantly adopting an
extended gaseous disk model (top panel of Fig. \ref{aitoffcirc}), except for the region at
the head of the Stream, which appears broader (bottom).

\section{Conclusions}

I carried out high resolution gravitational/hydrodynamical simulations of the interaction between the LMC and the MW using the orbital parameters suggested by the new HST proper motion measurements.
I find that ram-pressure stripping exerted by a tenuous MW hot halo during a single perigalactic passage forms  a Stream whose  extension and location in the Sky are comparable to the observed ones.
The stellar structure of the satellite is only marginally affected by tidal forces.\\

The numerical simulations were performed on the l SGI-Altix 3700 Bx2 at the University Observatory in Munich. This work was partly supported by the DFG Sonderforschungsbereich 375 ``Astro-Teilchenphysik''.


\begin{thebibliography}{}

\bibitem[Besla et al. (2007)]{Besla07} 
Besla, G., Kallivayalil, N., Hernquist, L., Robertson, B., Cox, T. J., van der Marel, R. P., 
\& Alcock, C. 2007, 
\textit{ApJ}, 668, 949 

\bibitem[Hernquist (1993)]{Hernquist93} 
Hernquist, L. 1993, 
\textit{ApJS}, 86, 389 


\bibitem[Kallivayalil et al. (2006a)]{Kallivayaliletal06a}
Kallivayalil, N., van der Marel, R. P., Alcock, C., Axelrod, T., Cook, K. H.,
 Drake, \& A. J., Geha, M. 2006, 
\textit{ApJ}, 638, 772

\bibitem[Kallivayalil et al. (2006b)]{Kallivayalil06b} 
Kallivayalil, N., van der Marel, R. P., \& Alcock, C. 2006, 
\textit{ApJ}, 652, 1213 

\bibitem[Klypin et al. (2002)]{Klypin02} 
Klypin, A., Zhao, H., \& Somerville, R. S.  2002, 
\textit{ApJ}, 573, 597 

\bibitem[Kroupa \& Bastian (1997)]{Kroupa97} 
Kroupa, P., \& Bastian, U. 1997, 
\textit{New Astronomy}, 2, 77 


\bibitem[Mastropietro et al.(2005)]{Mastropietro05} 
Mastropietro, C., Moore, B., Mayer, L., Wadsley, J., \& Stadel,
J. 2005, 
\textit{MNRAS}, 363, 509 


\bibitem[Mayer et al. (2006)]{Mayer06} 
Mayer, L., Mastropietro, C., Wadsley, J., Stadel, J., \& Moore,
B. 2006, 
\textit{MNRAS}, 369, 1021 



\bibitem[Piatek et al. (2007)]{Piatek07} 
Piatek, S., Pryor, C., \& Olszewski, E. W. 2007, 
\textit{ArXiv e-prints}, arXiv:0712.1764 
\textit{ArXiv e-prints},
\bibitem[Ruzicka et al. (2008)]{Ruzicka08} 
Ruzicka, A., Theis, C., \& Palous, J.\ 2008, 
\textit{ArXiv e-prints}, arXiv:0810.0968 


\bibitem[van der Marel et al. (2002)]{vanderMareletal02} 
van der Marel R. P., Alves D. R., Hardy E., \& Suntzeff N. B. 2002, 
\textit{AJ}, 124, 2639

\bibitem[Wadsley et al. (2004)]{Wadsley04} 
Wadsley, J. W., Stadel, J., \& Quinn, T.  2004, 
\textit{New Astronomy}, 9, 137 


\end{thebibliography}
\end{document}